\documentclass[a4paper,fleqn,usenatbib]{mnras}

\usepackage{newtxtext,newtxmath}

\usepackage[T1]{fontenc}
\usepackage{ae,aecompl}

\usepackage{graphicx}	
\usepackage{amsmath}	
\usepackage{amssymb}	
\usepackage[normalem]{ulem}

\title[Differential Least Squares Deconvolution]{A differential Least Squares Deconvolution method for high precision
spectroscopy of stars and exoplanets I. Application to obliquity measurements of HARPS observations of HD189733b}

\author[John B. P. Strachan, Guillem Anglada-Escude.]{
John B. P. Strachan,$^{1}$\thanks{E-mail: j.b.p.strachan@qmul.ac.uk }
Guillem Anglada-Escud\'e.$^{1}$
\\
$^{1}$School of Physics and Astronomy, Queen Mary University of London, 327 Mile End Rd., London, E1 4NS, UK\\
}

\date{Accepted XXX. Received YYY; in original form ZZZ}

\pubyear{2015}

\begin{document}
\label{firstpage}
\pagerange{\pageref{firstpage}--\pageref{lastpage}}
\maketitle

\begin{abstract}

High precision measurements of stellar spectroscopic line profiles and their changes over time contain very valuable information about the physics of the stellar photosphere (stellar activity) and can be used to characterize extrasolar planets via the Rossiter-McLaughlin effect or from reflected light from the planet.

In this paper we present a new method for measuring small changes in the mean line profile of a spectrum by performing what we call differential Least Squares Deconvolution (dLSD). The method consists in finding the convolution function (or kernel) required to transform a high signal-to-noise ratio template of the star into each observed spectrum. Compared to similar techniques, the method presented here does not require any assumptions on the template spectrum (eg. no line-list or cross-correlation mask required). 

We show that our implementation of dLSD is able to perform -at least- as good as other techniques by applying it to star-planet obliquity measurements of exoplanet  HD183799 during its transit. Among other things, the method should enable model independent detection of  light reflected by an exoplanet.

\end{abstract}

\begin{keywords}
techniques: spectroscopic -- planet-star interactions -- stars: rotation 
\end{keywords}



\section{Introduction}

Since the discovery of the first exoplanet 51 Peg b over 20 years ago using the Doppler method \citep{1995Natur.378..355M}  over 3500 exoplanets have been discovered 692 of which have been discovered with the Doppler method according to the exoplanet encyclopedia \citep{2011A&A...532A..79S}. The Doppler method indirectly infers the existence of an exoplanet from the radial velocity (RV) shifts in the spectra of the parent star caused by the reflex motion of the star due to the gravitational pull on it from the exoplanet.

Several high resolution spectrometers used for Doppler velocity measurements such as HARPS \citep{2000SPIE.4008..582P} operate in the visible wavelength range. More recently spectrometers such as CARMENES \citep{2010SPIE.7735E..13Q} have been built to operate in the near infra-red as well as the visible wavelength range and are in the process of carrying out radial velocity surveys on M dwarf stars in the solar neighbourhood.

If an exoplanet happens to transit its parent star then the Rossiter-McLaughlin (RM) effect,  (\cite{1924ApJ....60...15R}  and \cite{1924PA.....32..558M})  can be observed from high resolution spectra. The RM effect  was initially observed in spectra of eclipsing binary stars.  As the eclipsing star passed in front of the other rotating star the spectral lines were shifted due to asymmetry in blue shifted or red shifted light of the eclipsed star being blocked. The first RM effect observed for an exoplanet was observed over fifteen years ago for HD204958  (\cite{2000PASP..112.1421B} and \cite{2000A&A...359L..13Q}).

The RM effect depends on the projected spin orbit misalignment angle of the system and the projected rotational velocity of the star. The spin orbit misalignment angle is the angle projected on the sky between the rotation axis of the star and the normal to the orbital plane of the transiting exoplanet. Three different techniques have been used to determine these parameters for transiting exoplanet systems using high resolution spectroscopy.

The first method relies on retrieving these parameters based on the radial velocity measurements  e.g., \cite{2000PASP..112.1421B}, \cite{2000A&A...359L..13Q} and \cite{2009A&A...506..377T}. Systematic errors can occur in this method arising from the time-variable asymmetry of the stellar spectral lines during transit and solutions identified can be degenerate {\citep{2010MNRAS.403..151C}}.

Least  Squares Deconvolution (LSD) was introduced by \cite{1997MNRAS.291..658D} in order to detect magnetic fields in stars using spectropolarimetric observations. Observing the magnetic field signature in the spectropolarimetric observations was not possible to do directly due to the low signal to noise ratio (SNR) of the observation. LSD which involved deconvolving the observation with a template based on an atomic line list enabled the signal in each spectral line to be added and the resulting SNR to be increased by a factor of approximately the square route of the number of lines in an observation.  The disadvantages with this technique is that the list of atomic lines in the template has to be complete and their weight have to be known. 

 \cite{2010MNRAS.403..151C} introduced the use of the Cross Correlation Function (CCF) in order to track the shadow of the transiting exoplanet HD189733a as it passed across its parent star. The CCF is produced as described in \cite{1996A&AS..119..373B} and \cite{2002A&A...388..632P} using a template for the spectral type of the star which has a set of  box-shaped emission lines corresponding to the lines in the spectrum of the star. The template is then correlated with the spectrum for the star resulting in the CCF which is a single high SNR spectral line  which can be fitted to a Gaussian. As the exoplanet passes in front of the star the bump corresponding to the light blocked by the exoplanet can be seen moving across the CCF.   An alternative method using CCFs described in \cite{2016A&A...588A.127C}  has the advantage that it does not assume a particular Gaussian function for the CCF due to the subtraction of out and in-transit CCFs.
Again disadvantages with these techniques include ensuring all lines are correctly identified in the template and also catering for blended lines both of which are problems for late type stars.

In order to cater for the limitations here of the above techniques a new method called differential Least Squares Deconvolution (dLSD) has been developed. Instead of building templates by identifying lines from atomic linelists or from the star itself this method uses a high SNR template which is just a combination of spectra from the star.
The method consists in finding the convolution kernel that needs to match the template to the observation in a least squares sense.  As the light blocked by the exoplanet as it passes in front of the star is to first order the same as the spectra of the star though, Doppler shifted and inverted, the Kernel function will contain a sharply peaked function representing the blocked light. The Kernel function can then be fitted to a forward model. 

The fitting procedure consists of 1) generating the expected spectrum using the template and a parameterized physical model, 2) applying the deconvolution procedure to generate a synthetic kernel, and 3) apply the same deconvolution procedure to the observed spectrum to generate the kernel of the observation and 4) compare the two using a Bayesian procedure. The parameterized physical model includes among other things a limb darkening function, planet-star radius ratio on the obliquity of the orbit and the impact parameter (distance in plane of sky of closest approach of the exoplanet to the centre of the parent star). In this paper, we will only assume two free parameters (projected spin-orbit misalignment angle $\lambda$ and projected stellar rotational velocity $v_{eq}sin i$) for simplicity because these are the only ones that cannot be determined with photometry.

We describe in Section \ref{sec:LSD algorithm} the details of the dLSD algorithm and the forward model used to retrieve the projected spin orbit misalignment angle and projected rotational velocity parameters for a given system. In Section \ref{sec:Performance} we report on the performance of the dLSD algorithm based on test data and then for HARPS observations for HD189733. Finally the conclusions for the paper are given in Section \ref{sec:Conclusions}.

\section{Description of the dLSD algorithm}\label{sec:LSD algorithm}

High resolution spectra from systems such as HARPS are first processed using the HARPS-TERRA software \citep{2012ApJS..200...15A} and the option to remove the blaze from the spectra is selected. The HARPS-TERRA software creates a  high signal to noise (SNR) template spectrum T and provides the observation spectra O which have been velocity shifted to take account of the earth's rotation and also the measured Doppler shift of the star. 

Each spectrum is itself composed of a number of diffraction orders N. Thus for instance T = $\{ T_1, T_2,.T_r,..,T_{N}\}$ where $T_r$ represents the rth diffraction order.

The observations and template are interpolated using bicubic splines so that they have the same sampling and are moved to velocity space where $v_i$ represents the velocity of the ith sampled element.

The spectra are processed order by order. For the rth order the residuals $R_r$ to be fitted using least squares are obtained by subtracting the convolved spectrum $C_r$ from the observation:
\begin{equation}
    R_r(v_i) = O_r(v_i) - C_r(v_i),
	\label{eq:residuals}
\end{equation}

\noindent where the convolved spectrum is the convolution of our kernel with the template:

\begin{equation}
    C_r(v_i) = K_r*T_r(v_i).
	\label{eq:convolved spectrum}
\end{equation}

 In order to obtain $ K_r$ a deconvolution has to be performed. Firstly we approximate $K_r$ to be a linear combination of n basis functions $\theta_j$ representing the signal due to the transiting exoplanet plus a Dirac delta function which represents the signal from the star:
 
\begin{equation}
    K_r(v_i) \approx \sum_{j=1}^{n} \alpha_j\theta_j(v_i) + \delta(v_i),
	\label{eq:kernel}
\end{equation} 

\noindent where $\alpha_j$ are free parameters to be fitted.
 Top hat functions are used for the basis functions which have the property $\theta_j(v_i)$ is 1 when i=j and 0 otherwise.

We define a set of basis spectrum functions $b_j(v_i)$ to be the convolution of the jth basis function with the template:

\begin{equation}
    b_j(v_i) = \theta_j*T_r(v_i) = \sum_{k=1}^{s}\theta_j(v_k)T_r(v_i-v_k),
	\label{eq:b function}
\end{equation} 

\noindent where s is the number of velocity elements in the template. 

The residuals can now be expressed in terms of the observation and template spectra and the spectrum basis functions as:

\begin{equation}
    R_r(v_i) =O_r(v_i) - T_r(v_i) - \sum_{j=0}^{n} \alpha_j b_j(v_i).
	\label{eq:R in terms of b function}
\end{equation} 

Normally the number of velocity elements in the template (s) is significantly greater than the number of elements in the kernel (n) and we have an over-determined system which we fit using the least squares $\chi^2$ method:

\begin{equation}
   \chi^2(\alpha_0,...,\alpha_{n-1}) = \sum_{i=1}^{s}\frac{R(v_i)^2} {\sigma_i^2},
	\label{eq:chi squared function}
\end{equation} 

\noindent where $\sigma_i$ is the error/weight for each velocity element. The weight we use assuming Poisson statistics is:

 \begin{equation}
   \sigma_i = \sqrt{T_r(v_i).}
	\label{eq:error weight}
\end{equation} 

The template value is used as opposed to the observation value due to the template  not containing outliers from noise and thus producing more reliable weight estimates.

To minimise the $\chi^2$ we differentiate it with respect to each $\alpha_j$ and set the resulting terms equal to zero. Rearranging the resulting equations we get a set of n simultaneous equations which in matrix form is:

\begin{equation}
  AK_r = u,
	\label{eq:matrix equation}
\end{equation}

\noindent where $K_r = \{\alpha_1,\alpha_2,...,\alpha_{n}\}$ are our kernel coefficients and where each element in the n-column vector u is:

\begin{equation}
  u[m] = \sum_{i=1}^{s}\frac{b_m(v_i)R_r(v_i)}  {\sigma_i^2} ,
	\label{eq:u column vector}
\end{equation} 
and each element in the n$\times$n matrix A is:

\begin{equation}
  A[m][j] = \sum_{i=1}^{s}\frac{b_m(v_i)b_j(v_i)}  {\sigma_i^2} ,
	\label{eq:u array vector}
\end{equation} 

Deconvolution is an example of a Fredholm integral equation of the first kind and the solution to these equations are well known to often be ill-conditioned \citep{574131414}. In addition our matrix A will be rank-deficient as we would expect that the majority of the elements in K will be effectively 0 (within the noise level). In order to deal with these issues we use Tikhonov regularisation (\cite{SolnIllPosedProblems}) where the solution to the set of simultaneous equations is taken to be the minimisation of the following functional composed of an accuracy term and a penalty term:

\begin{equation} \label{eq:dLSD Deconvolution function}					
M(u,\kappa) = inf_{K_r \in F}\{||AK_r-u||^2 + \kappa||IK_r||^2\},
\end{equation}
where I is the identify matrix, F is the domain of $K_r$ and $\kappa$ is a free parameter commonly called the Tikhonov parameter expressing the relative weight of the penalty term to the accuracy term. The Tikhonov parameter has to be carefully chosen as picking a value too large will result in over-smoothing of the solution and if its too small then the noise will end up swamping the signal

There are several methods which can be used to select the value of the Tikhonov parameter including: Discrepancy principle, L-Curve criterion, General Cross Validation (GCV) and Normalized Cumulative Periodogram (NCP)(\cite{HansenInverseProblems}. An implementation of these methods is available in Regtools \citep{Hansen94regularizationtools} and were tested.  The NCP was selected as it worked well and did not have the disadvantages inherent in the other methods. The Discrepancy Method required manual selection of a safety parameter and as such is not ideal to use for automated software. The choice of this parameter was also sensitive to the resulting value determined for the Tikhonov parameter. The L-Curve criterion resulted in a dramatic over-smoothed  solution due to the cross-over between dominance of the accuracy term and penalty term not being sharp. The GCV suffered from a known issue of sometimes under-smoothing the solution.

Once solutions for the kernel for all N differential orders of the residuals have been determined we then combine these kernels using a simple mean:

\begin{equation} \label{eq:Kernel final}					
K(v_i) = \sum_{r=1}^N \frac{K_r(v_i)}{N}.
\end{equation}

An additional detail to the algorithm was that a mask M($v_i$) was used to identify velocity elements $v_i$ which were to be included in the calculation of the $\chi^2$ statistic (having value 1) or not (having value 0).
The mask was initialised so all its elements had value 1.
After calculating the residual spectrum R($v_i)$ in equation (\ref{eq:residuals}) clipping is performed to a user configurable level. All clipped velocity elements have their corresponding mask entry set to 0. In order to avoid problems with telluric contamination a  telluric mask $T_M(v_i)$ was used and Doppler shifted to move it to the same frame as the observations. Any velocity elements $v_i$ coinciding with the position of telluric lines in the mask had their entry in mask M($v_i$) set to 0.

\subsection{Forward model}

Our aim is to determine the projected spin-orbit misalignment angle of the exoplanet orbit and the projected rotational velocity of the star. In order to do this these two parameters are free parameters in our forward model. The forward model uses these free parameters along with a number of system parameters derived from photometry to derive a model kernel. This model kernel can then be compared to the kernel obtained from the in transit observations in order to confirm the validity of the values of the free parameters used. The rest of this section describes how the forward model enables us to produce the model kernel. In order to keep the forward model as simple as possible our model does not cover the time when the exoplanet is partially transiting the star - we only model from the time of second point of contact to the time of third point of contact during the transit.

We assume the star follows a standard quadratic limb darking law:

\begin{equation} \label{eq:Limb darkening}					
I(\mu) = I(0)  (1-\epsilon_1(1-\mu)-\epsilon_2(1-\mu)^2),
\end{equation}

\noindent where $\epsilon_1$ and $\epsilon_2$ are the limb darkening coefficients and $\mu$ is the cosine of the angle between the direction of the centre of the star to the observer and the centre of the star to the transiting exoplanet. $\epsilon_1$  and $\epsilon_2$ are fixed value input parameters to our forward model.

Following \cite{2015AJ....150..197H} we take the spectrum of the star out of transit to be:

\begin{equation} \label{eq:Sroot}	
S_{ROOT}(v) = S*G(v),
\end{equation}

\noindent where S is the spectrum of the star which has not been broadened due to the rotation of the star and G is the rotationally broadening profile of the star.

During the transit the spectrum of the star is given by:

\begin{equation}  \label{eq:Srit}	
S_{RIT}(v) = S*(G-D)(v)
\end{equation}

\noindent where D represents the light (which is not rotationally broadened) blocked by the planet. Assuming quadratic limb darkening the analytical functions for G and D are given in the appendix of \cite{2015AJ....150..197H} and are not repeated here except to say that G = G(u,$\epsilon_1$,$\epsilon_2$) and D = D(u,$\epsilon_1$,$\epsilon_2,x_p,y_p)$ where u is the relative velocity shift the broadening kernel is measured at and $x_p,y_p$ is the current position of the planet in the plane of the sky.


The projected spin-orbit misalignment angle $\lambda$, the angle between the angular momentum vector of the exoplanet and the axis of rotation vector of the star and the projected rotational velocity of the star $v_{eq}$sinI where $v_{eq}$ is the equatorial velocity of the star and I is the inclination of the spin axis with respect to the observer are the two free parameters in our forward model.

Here we specify two forward models to determine $x_p, y_p$ and $u$ from the free parameters specified in the previous paragraph and a number of fixed parameters derived from observables.

For the first model we assume that we do not  rely on the orbital parameters for the system. This can occur when only have old or partial radial velocity data for the orbit. In this case we rely on parameters derived from photometry including the second and third times of contact $t_2$ and $t_3$, the ratio of the exoplanet radius to that of the star $R_P/R_*$ and impact parameter b. The impact parameter is the minimum distance in the plane of the sky from the exoplanet to the centre of the star in units of stellar radius. For a transiting exoplanet b has a value between 0 and 1.

\begin{figure}
	\includegraphics[width=\columnwidth]{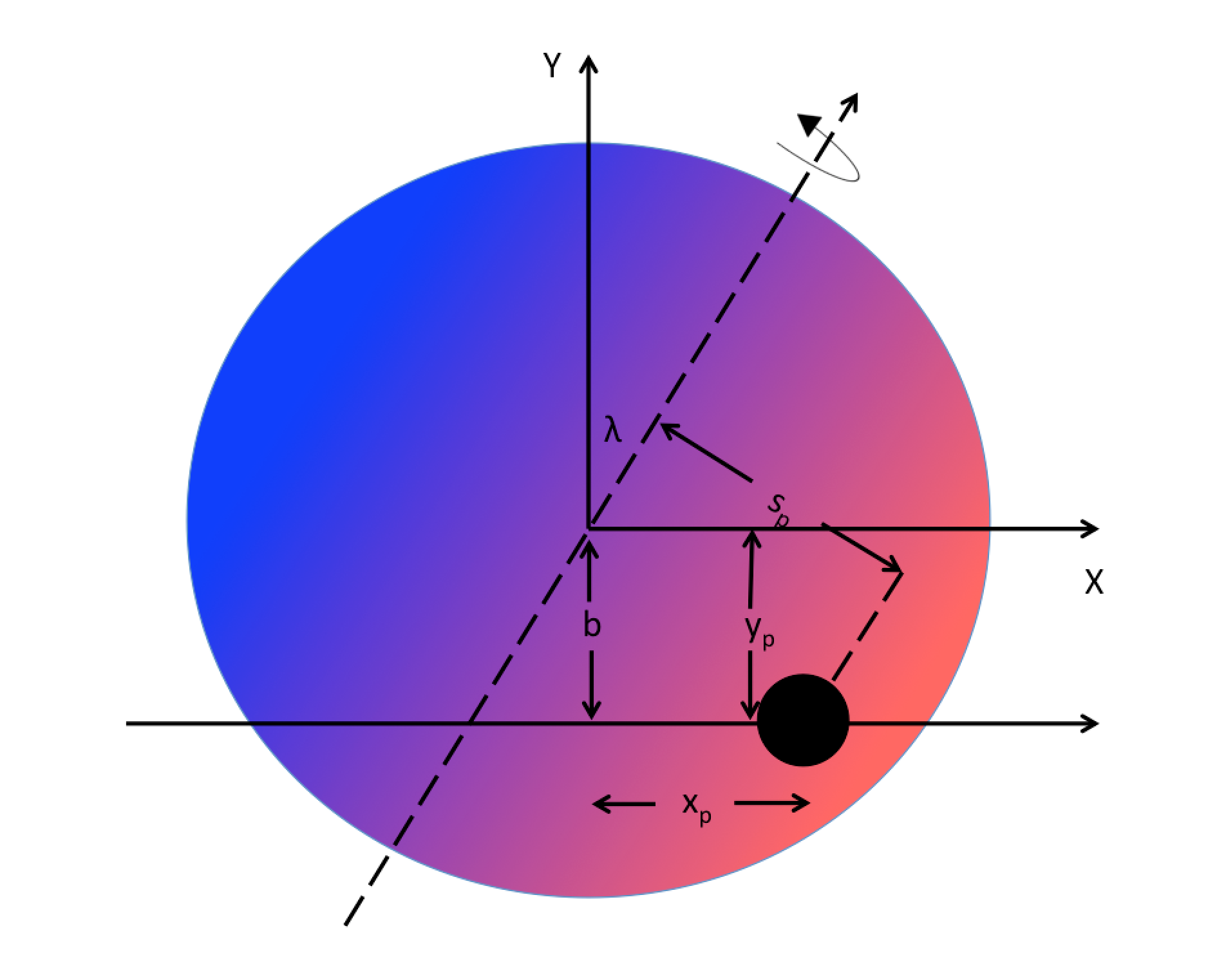}
    \caption{Planet during transit in the plane of the sky showing the spin-orbit misalignment angle $\lambda$ and impact parameter b. The path of the planet is also shown along with the axis used to locate the position ($x_p,y_p$) of the planet.}
    \label{fig:Basic planes of sky figure}
\end{figure}

For this case we assume circular motion of the exoplanet is valid and the exoplanet moves across the star in a straight line at a constant velocity $v_{planet}$.  The position of the exoplanet in the plane of the sky at any point in time is given by coordinate pair $(x_p, y_p)$ as detailed in Figure \ref{fig:Basic planes of sky figure}  Both of the axis are scaled so that $R_* = 1$. 

\begin{figure}
	\includegraphics[width=\columnwidth]{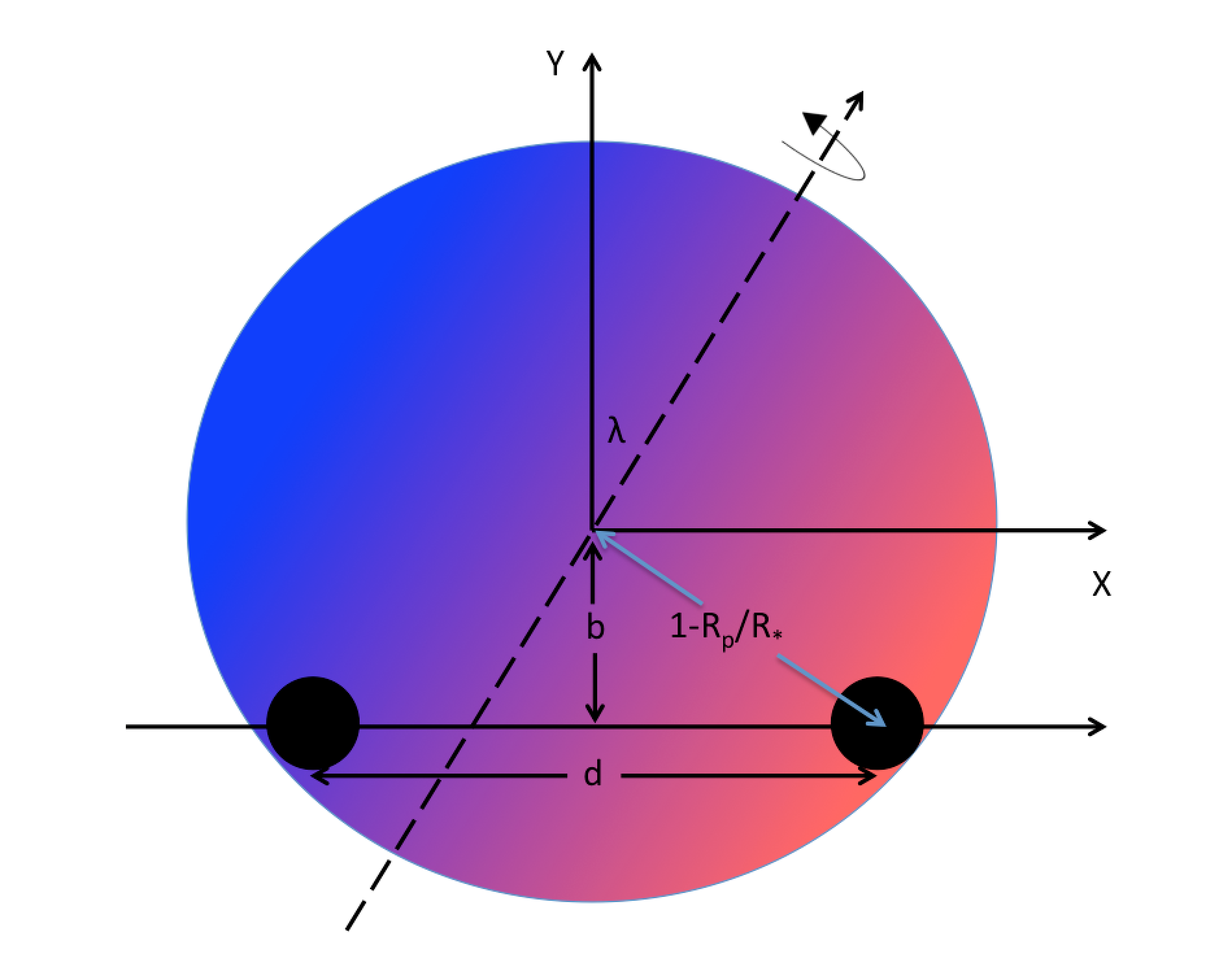}
    \caption{Plane of sky showing position of planet transiting at second and third points of contact and the distance d between them.}
    \label{fig:Plane of sky with POCs}
\end{figure}

Given the assumptions above and values for the projected spin-orbit misalignment angle $\lambda$ and impact parameter b the path of the planet across the star is on the line with $y_p$ = b. The distance between the second and third points of contact d in units of stellar radius (see Figure \ref{fig:Plane of sky with POCs}) is given by:

\begin{equation} \label{eq:d}					
d=2\sqrt{(1 - R_P/R_*)^2 - b^2}.
\end{equation}

Providing we know the times of second and third point of contact $t_2$ and $t_3$ from observations then we have that the velocity of the planet is:

\begin{equation} \label{eq:vplanet}					
v_{planet} = \frac{d}{t_3-t_2} = \frac{2\sqrt{(1 - R_P/R_*)^2 - b^2}}{t_3-t_2}.
\end{equation}
 
Thus the projected on the sky x-coordinate for the planet at time t is:

\begin{equation} \label{eq:x projected no orbit known}					
x_p(t) = v_{planet}(t-t_2) - \sqrt{(1 - R_P/R_*)^2 - b^2}.
\end{equation}

The shortest distance from the spin axis to the planet  $s_p$ again scaled in units of $R_*$  expressed in terms of $x_p$ and $y_p$ is using simple geometry from Figure \ref{fig:Basic planes of sky figure} :

\begin{equation} \label{eq:spin orbit distance}					
s_p(t) =x_pcos\lambda - y_psin\lambda. 
\end{equation}

The Doppler shift velocity u is:

\begin{equation} \label{eq:Doppler shift u}					
u(t)=s_pv_{eq}sinI,
\end{equation}

\noindent where v$_{eq}$ is the observed equatorial rotation velocity of the planet and I is the angle the rotation axis is inclined with respect to the observer.

For the case of the forward model if the orbit of the planet is known then the on sky coordinates are given by:

\begin{equation} \label{eq:x orbit known}					
x_p =rsin(\nu + \omega - \pi/2),
\end{equation}

\begin{equation} \label{eq:y orbit known}					
y_p =rcos(\nu + \omega - \pi/2)cosi,
\end{equation}
 
\noindent where $\nu$ is the true anomaly, $\omega$ is the argument of pericentre and i is the inclination of the orbit to the line of sight of the observer. We can then use equations \ref{eq:spin orbit distance}  and \ref{eq:Doppler shift u} to calculate the Doppler shift u.

For the purposes of the forward model we determine the non-rotationally broadened spectrum of the star S from equation \ref{eq:Sroot} by deconvolving the high SNR template T we have for the star with the rotational broadening profile G:

\begin{equation} 
S_{ROOT,r}(v) = T_r = S_r*G(v),
\end{equation}
where the suffixes r are present as we perform the deconvolution order by order. 

Having determined $S_r$ we can then determine $S_{RIT,r}$(v) from equation \ref{eq:Srit} by performing a convolution for each order r:

\begin{equation} 
S_{RIT,r}(v) = S_r*(G - D)(v),
\end{equation}

We can now determine the forward model kernel profile $M_r$ as specified in equation \ref{eq:dLSD Deconvolution function} using dLSD taking the template $T_r$ in equation \ref{eq:residuals} as $S_r$ and the observations $O_r$ in equation \ref{eq:convolved spectrum} as $S_{RIT,r}$(v).

\subsection{Bayesian model for parameter selection}

Bayes equation is used to determine the posterior  probability distributions $P(\theta|D)$ of the free parameters $\theta$ in the forward model based on the data D:

\begin{equation} \label{eq:Bayes equation}
P(\theta|D) = \frac{P(D|\theta)P(\theta)}{P(D)} \propto P(D|\theta)P(\theta),
\end{equation}

\noindent where $P(D| \theta)$ is the likelihood of the data given the parameters $\theta$, P($\theta$) is the prior probability distribution of the free parameters and P(D) is the normalization constant which we do not calculate here as only the value proportional to P($\theta|$D) is required to be determined.

Running dLSD on the spectroscopic data and the forward model synthetic data results in two files to be compared. We have the processed data file (D) and the forward model file (M) which depends on the free parameters $\theta$. Both files have the same format with each row containing a Kernel $k$ calculated for one of the spectra. Each row contains several elements $e$  which contains the value of the Kernel function at the velocity denoted by the element. Assuming Gaussian distributions for the uncertainties in the measurements the likelihood function L = P(D|$\theta$) is given by:

 \begin{equation} \label{eq:Overall likelihood}
L = \prod_ {k=1}^N \prod_ {e=1}^n \frac{1} {\sqrt{2 \pi (\sigma^2_ {k} + \sigma^2_ {ke} ) } }
exp\Big( -\frac{1}{2} \frac{ (D[k,e]- M[k,e])^2 }{\sigma^2_ {k} + \sigma^2_ {ke}  }      \Big),
\end{equation}

\noindent where N is the number of kernels, n is number of elements in a kernel, $\sigma_k$ and $\sigma_{ke}$ are the errors in the kernel and the kernel element respectively, and D[k,e] and M[k,e] correspond to the value of the eth element in the kth kernel for the data file and the model file respectively.  In order to avoid correlations between kernel element the kernels were sampled at the resolution of the spectroscope which for HARPS is 2.5kms$^{-1}$.  We tested that the noise in the kernel was Gaussian using test generated data samples with the Smirnof-Kolmogorov test.

We measured the noise for the kernel elements using the standard deviation of the kernel elements in the wings of the kernel where there is no signal. 

There are two parameters in the model: the observed rotation velocity of the star $v_{eq}$sinI and the projected spin orbit misalignment angle $\lambda$. We use flat priors for both of these parameters:

\begin{equation} \label{eq:Prior for vsinI}
P(v_{eq}sinI) = \frac{1}{ (v_{eq}sinI)_{max} - (v_{eq}sinI)_{min} },
\end{equation}

\noindent where $  v_{eq}sinI \in \{(v_{eq}sinI)_{min},...,(v_{eq}sinI)_{max}\}$.

\begin{equation} \label{eq:Prior for lamba}
P(\lambda) = \frac{1}{4\pi}, ~where  \lambda \in \{-2\pi,...,2\pi \},
\end{equation}

\noindent and where $(v_{eq}sinI)_{min}$ and $(v_{eq}sinI)_{max}$ represents the extremes in the range of values for $v_{eq}sinI$ and depends on the star.

In practice logarithms of values are calculated numerically so from Bayes equation the posterior probability of the free parameters $\theta$ becomes:

\begin{equation} \label{eq:log Bayes equation}
lnP(\theta|D)  \propto ln L + lnP(\theta),
\end{equation}

\noindent where logarithm of the likelihood lnL is:

\begin{multline} \label{eq:Overall loglikelihood}
lnL = nN ln \big( \frac{1}{\sqrt{2\pi} } \big) +
 \sum_ {k=1}^N \sum_ {e=1}^n ln \Bigg(  \frac{1} {\sqrt{(\sigma^2_ {k} + \sigma^2_ {ke} ) } } \Bigg)    \\ -  
\sum_ {k=1}^N \sum_{e=1}^n \frac{1}{2} \frac{ (D[k,e]- M[k,e])^2 }{(\sigma^2_ {k} + \sigma^2_ {ke} ) }.     
\end{multline}

Markov Chain Monte Carlo (MCMC) is used to calculate the posterior probabilities. The chain is started using an initial set of free  parameter values $\theta_0$ and lnLP($\theta_0$) is calculated.

A trial set of parameter is then calculated:

\begin{equation} \label{eq:Parameter gaussian perturbation}
\theta_{i+1} = \theta_i + step*\mathcal{N}(0,1),
\end{equation}

\noindent where $\mathcal{N}$(0,1) is a Gaussian distribution with mean 0 and variance 1. The step number is adjusted so that 20$\%$ of the trial set of parameters are accepted. The trial set parameters are accepted if: 

\begin{equation} \label{eq:trial acceptance condition 1}
lnLP(\theta_{i+1}) - lnLP(\theta_{i}) > 0,
\end{equation}

\noindent and if this condition is not met then the trial set of parameters can also be accepted if:

\begin{equation} \label{eq:trial acceptance condition 2}
Random[0,1] < exp[lnLP(\theta_{i+1}) - lnLP(\theta_{i}) ].
\end{equation}

The chains are normally allowed a burn in of 100 accepted proposals and are then allowed to run for 1000 accepted proposals.

\section{Performance}\label{sec:Performance}

\subsection{Testing by injecting simulated data}

\begin{figure}
	\includegraphics[width=\columnwidth]{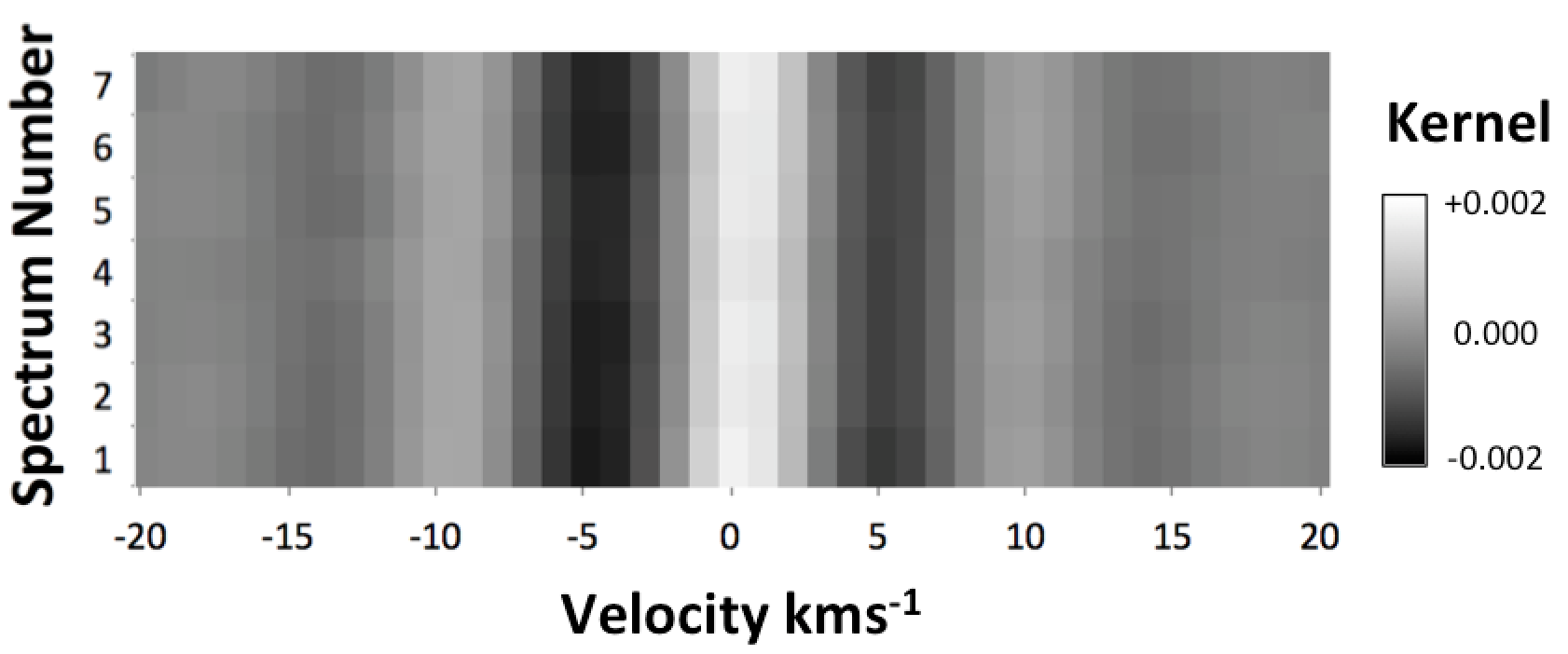}
    \caption{Greyscale of kernel function for seven out of transit spectra for HD189733 showing vertical striped bands near to the central velocities of the kernel.}
    \label{fig:OOT Kernel function}
\end{figure}

Testing of the algorithm was carried out using out of transit data from HARPS high resolution spectroscopic data for HD189733.    A high SNR template was created from the spectra captured that night and then the algorithm dLSD was run against seven out of transit spectra. The resulting kernels are shown in Figure \ref{fig:OOT Kernel function} showing a signal comprising of black and white vertical bends around the 0 velocity point of the kernel and which eventually disappears away from it. As flux is conserved in convolution operations the scale for the values of the kernel are such that if we had a kernel with all elements 0 bar one element with value -0.1 then this would correspond to the signal caused by the transiting planet having exactly the same spectrum as that of the star out of transit but with -1/10th of the amplitude.

This signal could be due to a number or reasons including correction for the blaze not being accurate enough and high frequency components in the spectra causing the banding due to aliasing from the deconvolution. As the signal is constant across the different Kernels it can be removed by averaging and then subtracting the average from each Kernel.

\begin{figure}
	\includegraphics[width=\columnwidth]{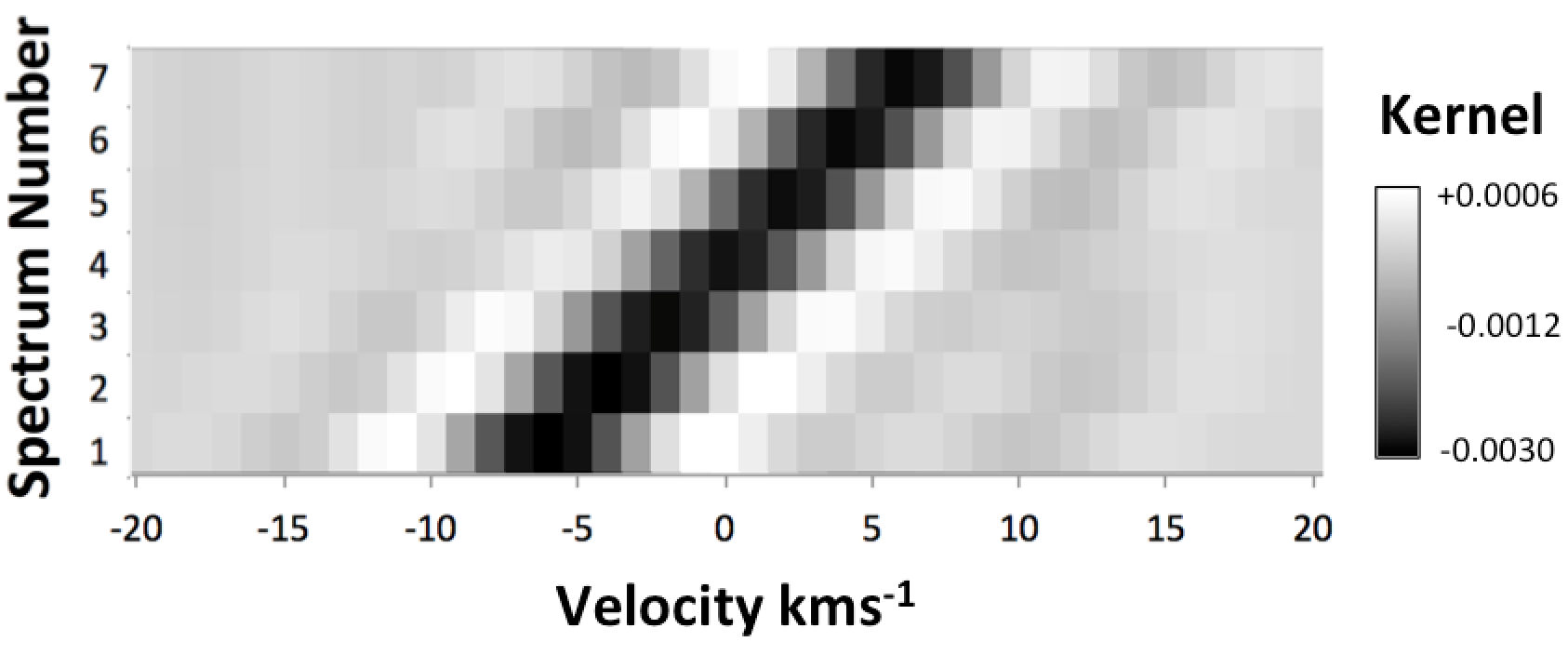}
    \caption{Greyscale of kernel functions for seven out of transit spectra for HD189733 with injected signal from -6 to +6 kms${^{-1}}$ in steps of 2kms$^{-1}$. Black diagonal line shows the presence of the signal along with alternative white and black banding caused by Gibbs effect.}
    \label{fig:Unnormalised Synthetic Kernel}
\end{figure}

The same seven spectra were used and then injected with a signal to simulate the effect of the planet transiting the star which took the form of adding a copy of the template spectrum whose amplitude was multiplied by a factor of -0.05 and whose velocity was shifted from -6 to +6 kms$^{-1}$ in steps of 2 kms$^{-1}$. The kernels for the seven spectra are shown in Figure \ref{fig:Unnormalised Synthetic Kernel} and show clearly the synthetic signal for the transiting planet - the black diagonal line moving from left to right. In addition there is ringing. The ringing has maximum amplitude of approximately 16\% of the amplitude of the main signal. The ringing is due to the Gibbs effect (\cite{Wilbraham1848})  which is well known to be present with Tikhonov regularisation where the kernel function is discontinuous or has sharp edges.

\begin{figure}
	\includegraphics[width=\columnwidth]{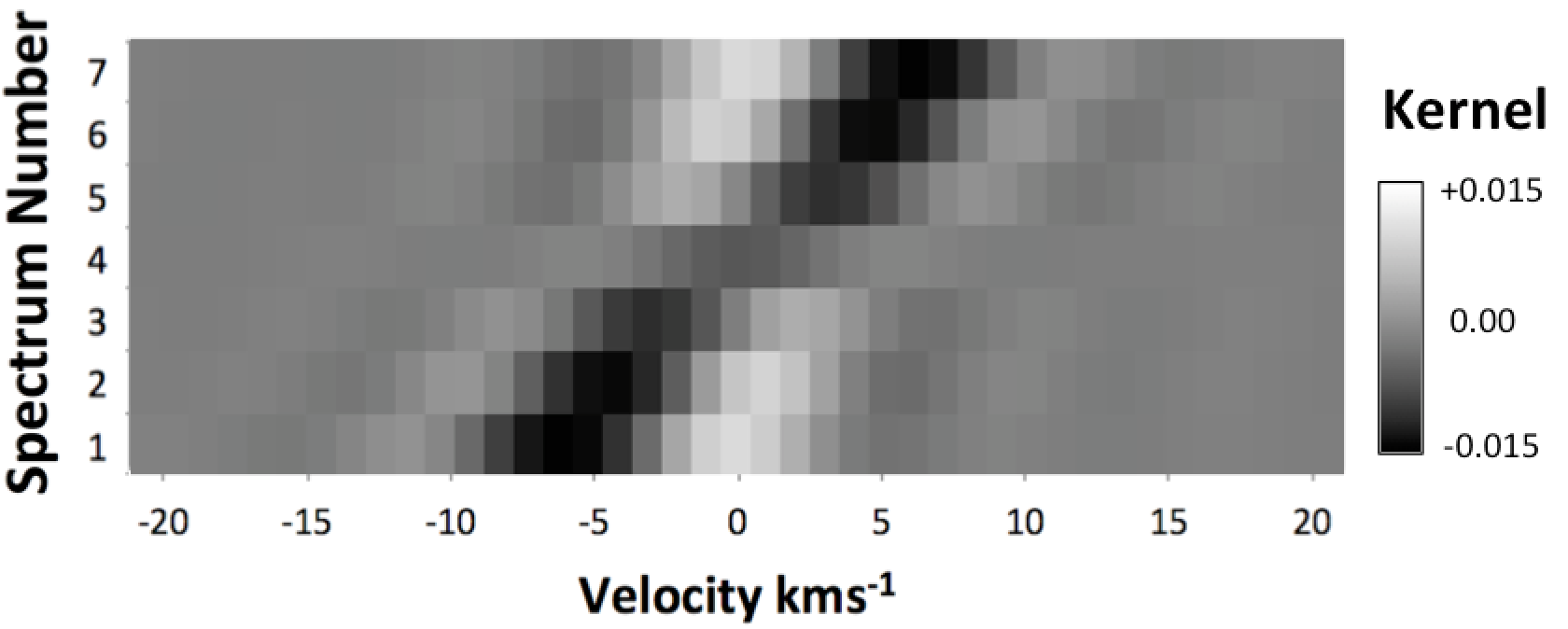}
    \caption{Greyscale of kernel function for seven out of transit spectra for HD189733 with injected signal from -6 to +6 kms${^{-1}}$ in steps of 2kms$^{-1}$ which has been normalised.}
    \label{fig:Normalised Synthetic Kernel}
\end{figure}

In the above the seven observation spectra were not normalised which will be the case for non-synthetic data. Normalizing these spectra gives the results for the kernel in Figure  \ref{fig:Normalised Synthetic Kernel} where at mid-transit the signal as weaker and is stronger towards the start and end of the transit.

\subsection{HD189733b}

HD189733b is a hot Jupiter transiting its parent star (\cite{2005A&A...444L..15B}) and is one of the most studied exoplanets. High resolution spectra of HD189733 were obtained using HARPS (High Accuracy Radial velocity Planet Searcher) at the 3.6 metre telescope in La Silla, Chile over four nights: July 30th, August 4th and September 8th 2006, and  August 29th 2007 under the allocated programme 079.C-0828(A). These spectra are publically available and were obtained by us from the ESO archive.

\begin{figure}
	\includegraphics[width=\columnwidth]{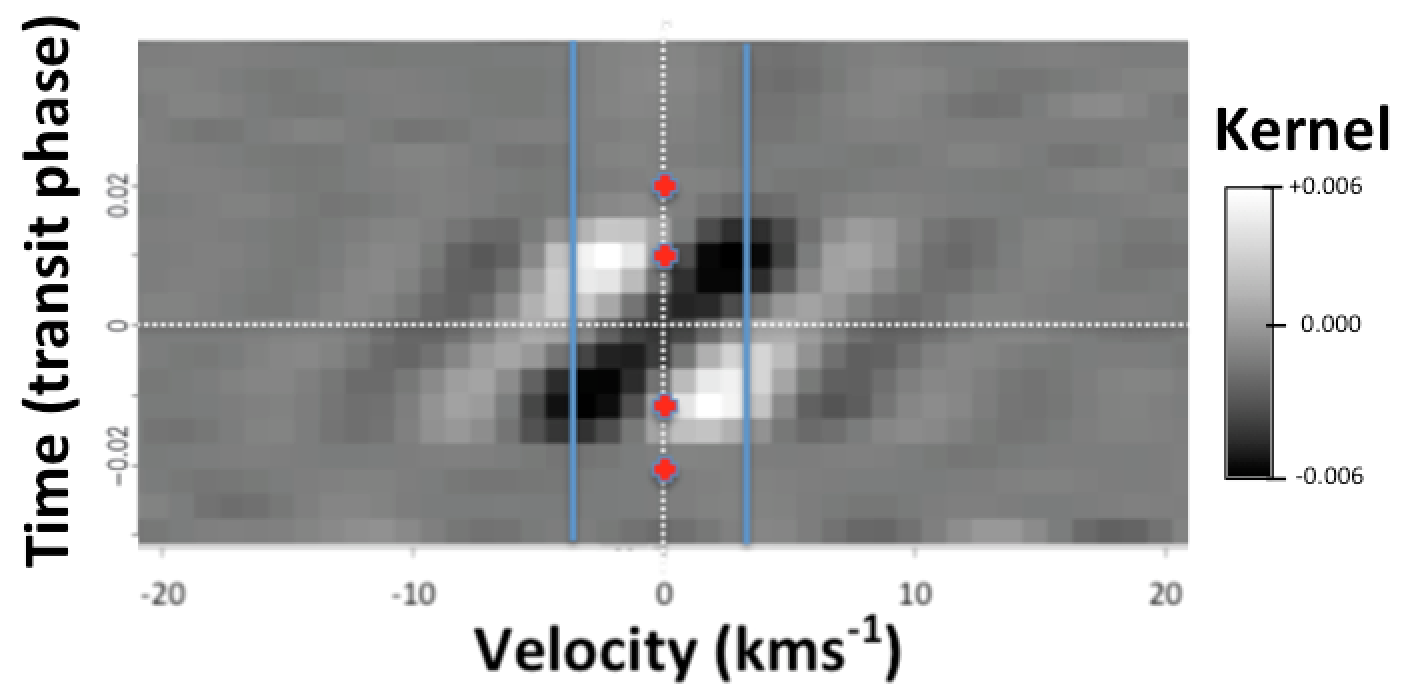}
    \caption{Greyscale of kernel function for the 20 spectra of HD189733 on the night of 9th August 2006. The lines in blue represent v$_{eq}$sinI determined for the star and the red dots represent the four points of contact during the transit.}
    \label{fig:HD189733 kernel function}
\end{figure}

The spectra were processed using HARPS-TERRA and then the dLSD software and the kernel for the spectra for the night of August 4th 2006 are shown in Figure \ref{fig:HD189733 kernel function}.

\begin{table}
	\centering
	\caption{Parameters used from previous studies of HD189733b.}
	\label{tab:standing data for HD189733}
	\begin{tabular}{lccr} 
		\hline
		Parameter & Value & Units & Reference\\
		\hline
		T$_C$ & 2453988.80339 & BJD& Triaud et al. (2009)\\
		T$_4$ - T$_1$ & 1.827 & hours & Winn et al. (2007)\\
		T$_2$ - T$_1$  & 24.6 & mins& Winn et al. (2007)\\
		b  & 0.687 & $R_*$ & Triaud et al. (2009)\\
		R$_p$/$R_*$  & 0.1581 &  & Triaud et al. (2009)\\
		\hline
	\end{tabular}
\end{table}

We used the simple forward model not relying on all the orbital parameters to determine the projected spin orbit misalignment angle and rotation of the star.  This was due to the radial velocities (RVs) available from HARPS not having good coverage of the complete orbit as the data came mainly from around the transit times.

HARPS-TERRA was used to derive the RVs from the HARPS data and a straight line was fitted to the out of transit measurements in order to determine the RVs of the star in transit.

The values we used for all the fixed parameters are detailed in Table \ref{tab:standing data for HD189733} with references to the literature where the values came from. There were two main sources for the data. The paper from \cite{2009A&A...506..377T}  who derived the parameters from both spectroscopic and photometric data. This paper had all the fixed parameters required for the forward model bar times for the second third and contacts. In order to obtain these times we took photometrically derived data from \cite{2007AJ....133.1828W}.

\begin{figure}
	\includegraphics[width=\columnwidth]{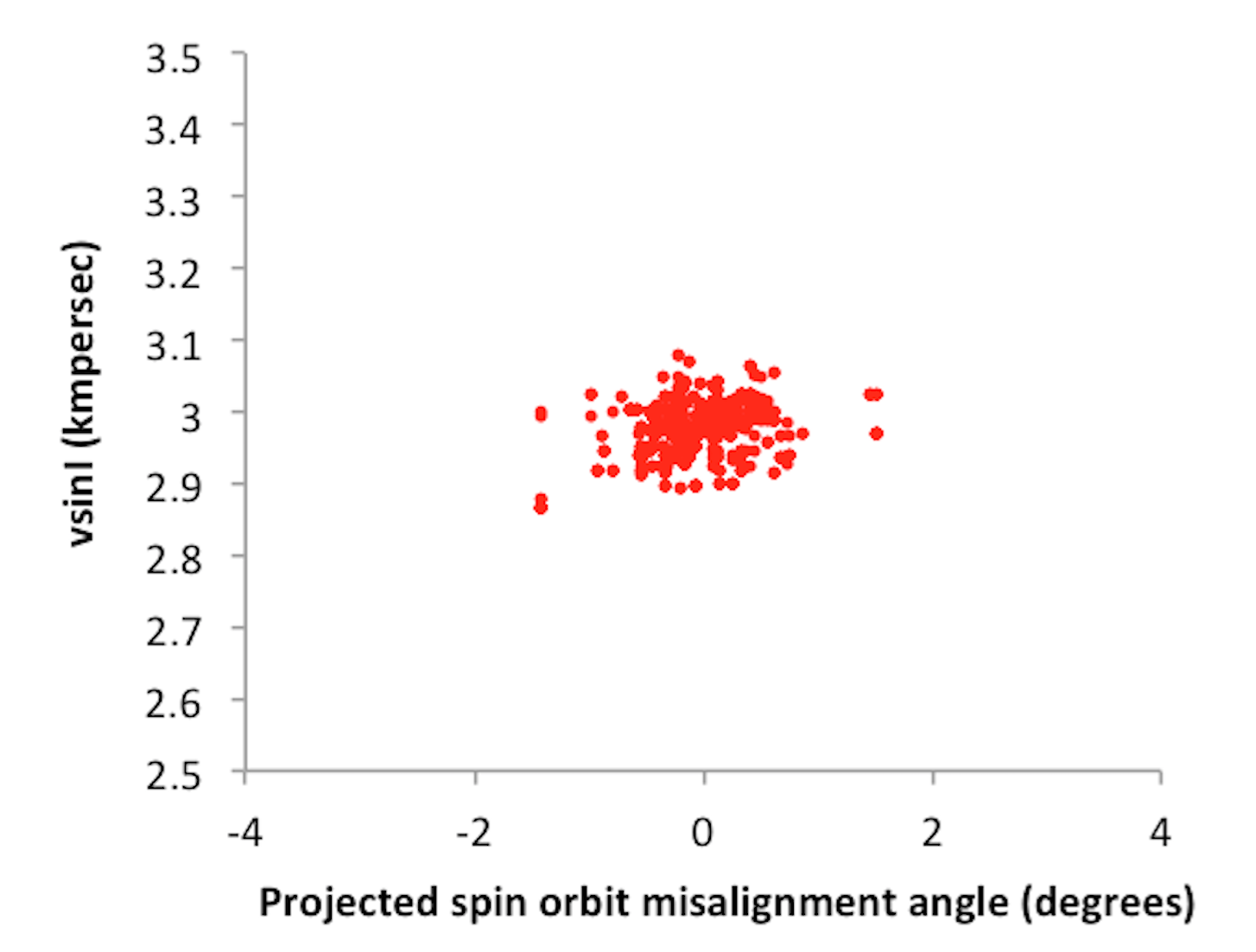}
    \caption{Correlation diagram showing the probability distributions for the parameters $\lambda$ and v$_{eq}$sinI for the transit of 8th September 2008 of HD189733. }
    \label{fig:correlation}
\end{figure}

In order to fit the model kernel to the kernel derived from the observations with our free parameters we use Markov Chain Monte Carlo (MCMC) with the Metropolis Hastings algorithm (\cite{1953JChPh..21.1087M}) The standard $\chi^2$ statistic was used to measure the fit. We allowed a burn of 100 accepted proposals before capturing a chain of 1000 accepted proposals. A correlation diagram for the probability distributions of the two free parameters  $\lambda$ and $v_{eq}sinI$ is shown in Figure \ref{fig:correlation}. This figure shows no significant correlation between the parameters.

\begin{table}
	\centering
	\caption{Mean and one sigma errors for the fitted parameters for the 7th September 2006 HD189733 data for differing values of the limb darkening parameters $\epsilon$.}
\label{tab:results}
	\begin{tabular}{lccr} 
		\hline
		Parameter & Value & Error & units\\
		\hline		
		$\epsilon_1$ = 0.7, $\epsilon_2 = 0.0$ & & & \\
		$\lambda$ & -0.374 & 0.475 & degrees\\
		$v_{eq}sinI$ & 2.988 & 0.015 & kms$^{-1}$ \\
		$\epsilon_1$ = 0.8, $\epsilon_2 = 0.0$ & & & \\
		$\lambda$ & -0.108 & 0.376 & degrees\\
		$v_{eq}sinI$ & 2.998 & 0.040 & kms$^{-1}$ \\
		$\epsilon_2$ = 0.9, $\epsilon_2 = 0.0$ & & & \\
		$\lambda$ & 0.007 & 0.416 & degrees\\
		$v_{eq}sinI$ & 3.021 & 0.043 &kms$^{-1}$ \\
		LDTK: $\epsilon_1$ = 0.7676 , $\epsilon_2 = 0.0028$ & & & \\
		$\lambda$ & -0.069 & 0.437 & degrees\\
		$v_{eq}sinI$ & 2.977 & 0.037 & kms$^{-1}$ \\
		\hline
	\end{tabular}
\end{table}

We produced chains for three different fixed values of the limb darkening parameter $\epsilon_1 = 0.7, 0.8$ and $0.9$ with $\epsilon_2$ held at 0 and the results are presented in Table \ref{tab:results}. The results show a small amount of correlation between the limb darkening parameter and the projected spin orbit misalignment angle.

We also calculated quadratic limb darkening parameters using the Limb Darkening Toolkit (LDTK) \citep{2015MNRAS.453.3821P} which uses the PHOENIX synthetic atmospheres and stellar spectra library \citep{2013A&A...553A...6H}. As inputs to the LDTK we used the HD189733 stellar parameters: effective temperature 4875 $\pm$ 43K,  logg 4.56 $\pm$0.03 and metallicity z = -0.03 $\pm$ 0.08 from \cite{2015MNRAS.447..846B}. The results of running dLSD against these limb darkening parameters are given in Table \ref{tab:results}.

We also ran Markov chains where instead of processing the spectra produced from the forward model with dLSD and comparing it with the dLSD output from the observation spectra we just compared the observation spectra directly with spectra generated from the forward model using a least squares fit. The Markov chains did not converge due to the lack of signal to noise in the RM effect using this method.

\section{Discussion and Conclusions}\label{sec:Conclusions}

We have developed and implemented the dLSD technique. The spin-orbit misalignment we calculated is within one $\sigma$ of  perfect alignment and is not far from previous published results such as the work of \cite{2009A&A...506..377T} who calculated $\lambda =0.87^{+0.32}_{-0.28}$ degrees and  \cite{2016A&A...588A.127C}  who calculated $\lambda =-0.4 \pm 0.2$ degrees.  The projected rotational velocity is in broad agreement with previous results whose values range from 2.9 to 3.5 kms$^{-1}$.

The rotation period of the star carried out photometrically by \cite{2008AJ....135...68H} is 11.953$\pm$ 0.009 days which if we assume the rotation angle is aligned with us gives sinI is 1 and using an average of the three values we obtained for $v_{eq}$sinI gives us a radius of 0.714R$_{\odot}$. This value is significantly smaller than the values calculated by \cite{2009A&A...506..377T} of R$_*$=0.766$\pm^{+0.007}_{-0.013} R_\odot$ and of the R$_*$=0.805$\pm 0.016R_\odot$ value using interferometry from \cite{2015MNRAS.447..846B}. However the discrepancy may be due to the differential rotation of the star as explained in \cite{2010MNRAS.403..151C}.

One current drawback with the dLSD approach is the amount of computer time to perform the calculation in the forward model. 

For the analysis of HD189733 to calculate one step in the MCMC chain takes ~5 minutes on a laptop with a 2.3GHz I7 processor. 

Given that only $20\%$ of the steps are accepted the time taken to perform all the accepted steps including the burn in is just over 5 days. However there is significant room to decrease the time of the individual steps by introducing parallel processing into the algorithm so that each of the spectral orders could be processed separately.

In future work the authors expect to be able to draw quantitative comparisons between the different methods CCFs, LSD and dLSD discussed here for different types of stars.

In conclusion we have shown that we have been able to build the dLSD algorithm which can successfully be used to detect the path of an exoplanet as it transits its parent star using high resolution spectroscopy. In particular we have shown that the only template we need to use is one which is just built directly from the spectra of the star itself and from this we have been able to estimate the spin orbit misalignment angle for  the HD189733 system.

\section*{Acknowledgements}

J.S. acknowledges support of studentship funded by Queen Mary University of London. This research was also supported by STFC Consolidated Grant ST/P000592/1.



\bibliographystyle{mnras}
\bibliography{bib-various} 









\bsp	
\label{lastpage}
\end{document}